\documentclass[preprint,a4paper,superscriptaddress,floatfix]{revtex4-2}

\usepackage{times}
\usepackage{graphicx,xcolor}
\usepackage{amsmath,amsfonts,amssymb,bm}
\usepackage{listings}
\usepackage{soul}
\usepackage{multirow}
\usepackage{makecell}

\usepackage[
unicode=true,
bookmarks=true,%
bookmarksnumbered=true,%
hyperfootnotes=false,
colorlinks=true,%
linkcolor=blue,%
citecolor=blue,%
urlcolor=blue
]{hyperref}  

\DeclareMathOperator{\sech}{sech}

\newenvironment{mat}{\left[\begin{array}{ccccccccccccccc}}{\end{array}\right]}
\newcommand\bcm{\begin{mat}}
\newcommand\ecm{\end{mat}}

\newenvironment{cmat}{\left(\begin{array}{ccccccccccccccc}}{\end{array}\right)}
\newcommand\bcrm{\begin{cmat}}
\newcommand\ecrm{\end{cmat}}

\newenvironment{rmat}{\left[\begin{array}{rrrrrrrrrrrrr}}{\end{array}\right]}
\newcommand\brm{\begin{rmat}}
\newcommand\erm{\end{rmat}}

\begin{document}

\title{Unveiling Solitonic Collisions in Mechanical Metamaterials}

\author{Yasuhiro Miyazawa}
\affiliation{Department of Mechanical Engineering, Seoul National University, 1 Gwanak-ro, Gwanak-gu, Seoul 08826, South Korea}
\affiliation{Department of Aeronautics and Astronautics, University of Washington, Seattle, Washington 98195}
\author{Christopher Chong}
\affiliation{Department of Mathematics, Bowdoin College, Brunswick, Maine 04011}
\author{Panayotis G. Kevrekidis}
\affiliation{Department of Mathematics and Statistics, University of Massachusetts Amherst, Massachusetts 01003-4515}
\affiliation{Department of Physics, University of Massachusetts Amherst,  Massachusetts 01003}
\author{Jinkyu Yang}
\email[]{jkyang11@snu.ac.kr}
\affiliation{Department of Mechanical Engineering, Seoul National University, 1 Gwanak-ro, Gwanak-gu, Seoul 08826, South Korea}

\date{\today}
\begin{abstract}
Interactions between solitary waves are crucial for understanding nonlinear phenomena in systems such as optics, fluid dynamics, and mechanical metamaterials. Rarefaction solitary waves, in particular, offer insight into nonlinear wave dynamics in strain-softening media. Despite their proposed applications in waveguides and energy harvesting, key characteristics—particularly solitonic collisions—are not yet fully understood due to energy dissipation and the need for high-precision measurement techniques. In this work, we introduce an experimental platform for studying pure rarefaction solitons in a strain-softening lattice. 
Our results show that both symmetric and asymmetric collisions display elastic interactions and amplitude-dependent phase shifts. The experimentally observed dynamics, including soliton speed and phase shifts, closely match numerical simulations and analytical predictions based on the Boussinesq approximation. These findings not only validate our platform but also highlight the potential of mechanical rarefaction solitons for probing nonlinear wave interactions and advancing wave-based computing.
\end{abstract}

\maketitle


\section{Introduction}
Studies of solitary waves---characterized by their stable propagation of localized energy---have long served as a foundational framework for understanding nonlinear phenomena across diverse disciplines, such as plasma~\cite{Liang2009, Han2008, Xu2011, NarayanGhosh2012, Alam2021, Shohaib2022}, fluid~\cite{Matsuno1994, Craig2006}, optics~\cite{Mamyshev1991, Bigo1997, Li2018c, Hasegawa2022, Zhou2023}, as well as condensed matter physics~\cite{Billam2011, Nguyen2014}.
One of the most notable properties of solitons---and quintessential to their very definition---is their collision dynamics, where waves retain their shape after interaction, and merely exhibit a phase shift~\cite{Ablowitz2011a}.
These unique soliton interactions have been extensively explored in the aforementioned disciplines, often through mathematical techniques such as the inverse scattering transform (IST)~\cite{Kivshar1989, Ablowitz2011a}, the Hirota method~\cite{Yu2019a, Liu2019, Zhou2023, Shohaib2022}, perturbation methods~\cite{Kivshar1989}, asymptotic analysis~\cite{Ablowitz2011a}, and the Poincaré-Lighthill-Kuo (PLK) method~\cite{Xu2011}, among others.
These tools have been essential in deriving soliton solutions and expanding the understanding of the rich interaction dynamics thereof.

A study of solitary wave interaction dynamics not only enhances fundamental knowledge about nonlinear waves but also opens up a plethora of application avenues across various scientific and technological disciplines.
For instance, in optics, it has been reported that controlling interaction and phase shift can enhance the signal propagation distance within optical fibers~\cite{Liu2019}.
Another proposition for such practical use of soliton interaction can be inferred from a growing interest in leveraging solitons for computational purposes.
The concept of using solitons to transmit information and perform computations through their collisions is being investigated as an alternative to traditional computing paradigms~\cite{Adamatzky2012, Zhou2022, Wang2023}.
Approaches such as all-optical logic gates~\cite{Zhou2022, Mukherjee2023, Wang2023}, implementations in reservoir computing~\cite{Maksymov2023}, and quantum solitons as qubits~\cite{Bullough2003} demonstrate that the unique properties of solitons may lead to novel computing technologies.

On another front, the emergence of mechanical metamaterials in recent decades has spurred renewed interest in nonlinear waves in engineered nonlinear media~\cite{Agranovich2004, Denz2010, Poutrina2010, Manimala2018, Xu2019, Li2021c, Fronk2023, Fang2024}.
Among these, solitary wave propagation in mechanical systems has been a key area of study~\cite{Herbold2013a, Maurin2016, Maurin2016a, Chong2017, Kim2017, Yasuda2019, Khajehtourian2021, Ye2023}; see also the relevant books~\cite{Nesterenko2001, yuli_book, Chong2018} and reviews~\cite{SEN200821}.
Much of the focus has been primarily on compressional solitary waves in classical granular media~\cite{Zhen-Ying2007, Carretero-Gonzalez2009, Chong2017, Kim2017, Kekic2018, Liu2021a, Zhang2021d} that exhibit strain-hardening behavior.
Recent studies have extended attention to strain-softening metamaterials, including origami-based metamaterials~\cite{Yasuda2019}, bar-linkage lattices~\cite{Ye2023}, buckled beam~\cite{Maurin2016, Maurin2016a, Deng2019b}, and LEGO-based lattices~\cite{Deng2018}.
In such settings, so-called rarefaction solitary waves are possible. In a rarefaction wave, the tensile force is spatially localized, rather than the compression force as in a traditional solitary wave.
While these studies have elucidated the potential of solitary waves to be controlled in mechanical settings and used for robust waveguiding, impact mitigation, or energy harvesting applications, their primary focus has been the generation and propagation of solitary waves, which does not involve the dynamics of their interactions.
Quite remarkably, experimental studies of rarefactive solitary wave interactions in mechanical systems, including, notably, direct measurement of phase shifts, are still in their nascent stages.
This contrasts with experimental studies in non-mechanical settings, such as optics~\cite{Shou2011, Mitschke1987}, LC circuits~\cite{Islam1987}, and fluids~\cite{Guizien2001}, as well as atomic Bose-Einstein condensates~\cite{Nguyen2014}, where attempts have been made to measure interaction dynamics and phase shifts.
The lack of previous studies in mechanical metamaterials can be attributed to several factors involved in the experiments: generating pure soliton inputs, significant energy dissipation (during the propagation towards collision), a low signal-to-noise ratio in full-field measurements, and, perhaps most crucially, the key challenge of accurately measuring and detecting phase shifts in the millisecond range.
Furthermore, the methodology for processing soliton interactions and detecting phase shifts has yet to be established.
It is the scope of this work to provide a paradigm that will qualitatively and quantitatively overcome these constraints, paving the way for substantial 
further progress in the field.

To address these limitations, our study introduces an experimental platform for investigating the interactions of mechanical rarefaction solitary waves.
The platform consists of a one-dimensional mechanical metamaterial lattice with strain-softening nonlinearity and low energy dissipation, enabling the generation and precise measurement of phase shifts during wave collisions.
By employing a combination of full-field measurements and a robust post-processing methodology, we achieve high accuracy in detecting phase shifts, which are further validated against analytical and numerical predictions.
This work adds a key layer of experimental insights into phase shifts in rarefaction solitary wave collisions within mechanical systems, bridging theoretical models and experimental observations.
More importantly, the methods and results presented here could provide a prototypical framework for future studies, offering a foundation for exploring nonlinear dynamics and guiding the design of advanced metamaterials and other engineered systems.

The paper is organized as follows:
We first introduce our nonlinear mechanical metamaterial platform and cover the mathematical model in Sec.~\ref{sec:setup}.
Section~\ref{sec:approximation} further delves into the analytical approximate solutions of our nonlinear lattice at its continuum limit.
We then conduct a prototypical experiment and numerical simulation in Sec.~\ref{sec:onesoliton} to verify the analytical solution.
The verification is followed by the analytical, numerical, and experimental study on the head-on collision of two solitary waves in Sec.~\ref{sec:twosoliton}.
We derive an equation to predict the phase shift in the rarefaction solitary waves after collision, which is favorably compared against the numerical and experimental results.
Lastly, Section~\ref{sec:discussion} presents a brief summary and concludes our findings, along with the suggestion of a number of interesting future directions.

\section{Results}
\subsection{Physical setup and mathematical model}\label{sec:setup}

\begin{figure*}[t]
    \centering
    \includegraphics[width=\linewidth]{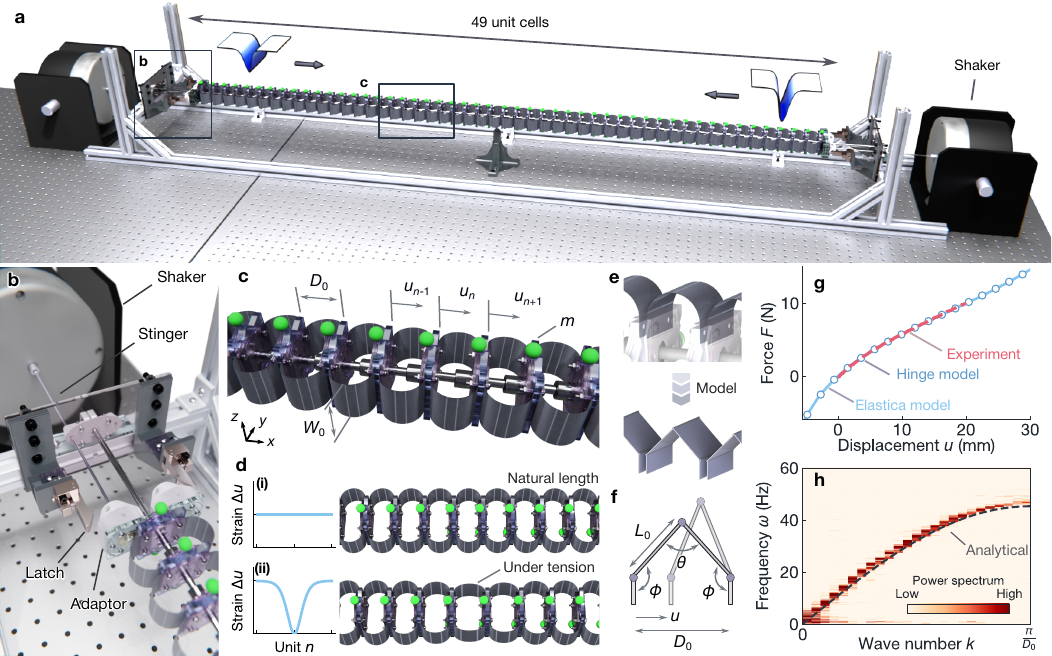}
    \caption{Physical setup and mechanical properties of 1D strain-softening lattice.
    (\textbf{a}) Overview of the lattice consisting of 49 unit cells with shakers attached on the boundaries for excitation.
    (\textbf{b}) Detailed view of the boundary structure connecting the lattice and shaker~(Br\"uel \& Kj{\ae}r modal exciter type 4825).
    (\textbf{c}) Overview of the lattice comprised of acrylic plates (i.e., rigid connectors with mass $m$, referred to as particles in this study), which connect thin sheets of spring steel that act as elastic springs under bending.
    The green spherical markers are used for motion tracking.
    (\textbf{d}) Lattice in (i) equilibrium and (ii) tension due to the rarefaction solitary wave.
    (\textbf{e}) Schematic illustration of the torsion spring modeling and
    (\textbf{f}) definition of geometry and variables of the model.
    (\textbf{g}) Force-displacement relationship of a unit cell.
    Blue solid line, the elastica model~(see Supplementary Note~1);
    blue open circle symbols, equivalent torsion spring model~[cf. Eq.~\eqref{eq:force_relation}];
    red dashed line with shaded region, uni-axial compression experiment result.
    The red dashed line and shaded region represent the mean and standard deviation of five trials, respectively.
    (\textbf{h}) Linear wave dispersion relationship.
    Black dashed line, infinite lattice dispersion relationship;
    surface map, 2D FFT spectrum of chirp excitation experimental results using 50-particle lattice.
    }
    \label{fig:schematic}
\end{figure*}

In the current study, we consider a one-dimensional~(1D) lattice composed of 49 unit cells with a longitudinal degree of freedom shown in Fig.~\ref{fig:schematic}a.
The lattice is connected to modal shakers on each boundary, which are used to generate solitary waves.
As shown in Fig.~\ref{fig:schematic}b, a connection between the shaker and lattice is made through a $3$ mm diameter carbon steel stinger, which is tightly fixed to the boundary particle with an adaptor.
When generating a rarefaction wave, the stingers are pulled outward, and the boundary particle is displaced and then held by a pair of latches to prevent any ``backlash'' compressive waves from being induced, leaving only tensile waves within the lattice~(details of latch operation will be discussed in the later section and in Supplementary Video~1).

Figure~\ref{fig:schematic}c shows a detailed view of the lattice, where a pair of two semi-cylindrical 0.1 mm-thick spring steel sheets are connected via rigid connectors on each end, which are $D_0=30.0$ mm apart from each other~(i.e., $D_0$ is the diameter of the hollow cylinder and a lattice constant).
The rigid connectors consist of four acrylic plates enclosing a ball-bearing, collectively treated as particles with mass $m$. 
For our setup with 49 unit cells, the lattice consists of 50 particles, including boundary particles.
A stainless steel shaft with a diameter of $6.35$ mm passes through the ball-bearing, restricting the lattice to axial movement only.
The displacement $u_n$ is defined as the displacement of the $n$-th particle.
The spring steel sheet has dimensions $W_0=30.0$ mm and a length of $\dfrac{\pi D_0}{2}\approx47.1$ mm, such that the lattice constant becomes $D_0$ mm when bent into a semi-cylinder.
Each particle is equipped with a pair of green markers to be tracked by a high-speed camera, enabling the full-field measurement of rarefaction solitary wave propagation.
(For more details on fabrication, assembly, and experimental setup, see the Methods section and Supplementary Figure~1.)

Under rarefactive excitation from the shaker, the unit cell undergoes tensile deformation, as depicted in Fig.~\ref{fig:schematic}d.
Here, the strain in the left two panels is defined as axial compressive strain $\Delta u_{n}=u_n-u_{n+1}$.
When the rarefactive wave is generated by the shaker and transmitted to the lattice via a stinger, the spring steel sheet becomes semi-elliptical under tension~[sub-panel~(ii)] from a semi-circular profile in natural length~[sub-panel~(i)]~(more details of the unit deformation can be found in Supplementary Figure~2 and Supplementary Video~2).

\subsubsection{Hinge model and equation of motion}
One way to describe the deformation and underlying mechanical behavior of a unit cell is through a so-called inextensible elastica model~\cite{Plaut2014, Virgin2018}. 
This is essentially a thin beam model that describes the force-displacement relationship and involves solving a simultaneous system of ordinary differential equations~(see Supplementary Note~1 and Supplementary Figure~2 for details).
While the inextensible elastica model accurately describes static behavior~\cite{Plaut2014, Virgin2018}, it is unsuitable for dynamic simulation due to its high computational cost that involves solving implicit force-displacement relationships at each time step.
Furthermore, the elastica model does not lend itself to the analytical approximations we present in the later sections.
Therefore, we employ a simplified model to study the dynamics of the unit cell hereafter, in which the force-displacement relationship is expressed explicitly. 
Specifically, we consider a unit cell model consisting of three hinges, as shown in Fig.~\ref{fig:schematic}e.
The kinematic behavior can directly be expressed in terms of the folding angle $\theta$ and auxiliary angles $\phi$~[defined in Fig.~\ref{fig:schematic}f], which are governed by the independent variable $\Delta u$, the axial strain of the unit, as follows:
\begin{align}
    \theta=2\sin^{-1}\left(\dfrac{D_0-\Delta u}{2L_0}\right),\quad\phi=\pi-\sin^{-1}\left(\dfrac{D_0-\Delta u}{2L_0}\right),
\end{align}
where $L_0=\sqrt{2}D_0/2\approx21.2$ mm is the hinge length~(i.e., radius of torque).
The illustration in Fig.~\ref{fig:schematic}f shows the deformation of the reduced model under displacement $u$ applied at the left hinge.
Moreover, the restoring force due to the torsion spring, where $\theta$ is defined, also serves to prevent asymmetric deformation~(i.e., unequal $\phi$'s).
Then, the dynamics of the lattice is characterized by the Lagrangian,
\begin{align}
    L=\sum_{n\in\mathbb{Z}}\left[\dfrac{1}{2}m\dot{u}_n^2-U_n(\Delta{u}_n)\right],
\end{align}
where $\mathbb{Z}$ is an integer indexing the particles, $u_n$ is the axial displacement of the $n$-th particle, and $\Delta{u}_n=u_n-u_{n+1}$ is the axial compressive strain of the $n$-th unit cell.
The potential energy $U_n$ of each unit cell is described as
\begin{align}
    U_n(\Delta{u}_n)&=\sum_{i=1}^3\dfrac{2W_0}{i+1}k_i\left[\left(\theta-\theta_0\right)^{(i+1)}+2\left(\phi-\phi_0\right)^{(i+1)}\right],
\end{align}
assuming a cubic nonlinear torsion spring, where $k_i$ is a torsion spring coefficient per unit length, and $\theta_0$ and $\phi_0$ are, respectively, the folding angle and auxiliary angle when the system is in equilibrium.
In the absence of pre-compression or tension, equilibrium angles can be determined as $\theta_0=\dfrac{\pi}{2}$ and $\phi_0=\dfrac{3\pi}{4}$.
Using the Euler-Lagrange prescription, the equation of motion for the $n$-th particle reads,
\begin{align}\label{eq:eom_hinge}
    m\ddot{u}_n+F\left(\Delta u_n\right)-F\left(\Delta u_{n-1}\right)+c_{\rm damp}\dot{u}_n=0.
\end{align}
Note that we added a phenomenological description of energy dissipation in the form of a simple dash-pot model $c_{\rm damp}\dot{u}_n$, in line with earlier studies~\cite{Nesterenko2001, Chong2018}, to describe the friction between the shaft and ball-bearing.
While more elaborate forms of dissipation have been proposed and analyzed~\cite{Carretero-Gonzalez2009, lindenberg, vergara, James_2021}, we have found the above form to be quantitatively sufficient for the purposes of our experimental configuration.
We use $c_{\rm damp}=0.248$ kg/s for this study~(more details regarding damping coefficient estimation can be found in Supplementary Note~2, Supplementary Figure~3, and Supplementary Video~3).

The unit cell internal forces are explicitly written as,
\begin{align} \label{eq:force_relation}
    F(\Delta u)&=\dfrac{-4}{\sqrt{4{L_0}^2-(D_0-\Delta u)^2}}\nonumber\\
    &\quad\quad\times\sum_{i=1}^{3}k_iW_0\left[2^i+\left(-1\right)^{i+1}\right]\left(\frac{\theta-\theta_0}{2}\right)^i,
\end{align}
The resultant force-displacement relationship exhibits strain-softening behavior, as shown in Fig.~\ref{fig:schematic}g, denoted as blue open circle symbols, which shows good agreement with the experimental result~(red dashed line).
We also note in passing that the force-displacement relationship from the hinge model matches the inextensible elastica model, denoted as the blue solid lines, despite its simplicity~(see Supplementary Note~3 and Supplementary Figure~4 for the details 
associated with the static behavior within
the experiment).

For numerical simulations, we solve Eq.~\eqref{eq:eom_hinge} using an 8th-order Dormand-Prince scheme with a time step size of $\Delta t=10^{-6}$ s and error tolerance of $10^{-12}$ to determine the time step adaptively.
There are $N=50$ particles in the simulation, where the first node~($n=-24$) is driven
(formula given in the next section), while we primarily use a fixed boundary condition for the rightmost node~($n=25$).
When simulating solitary wave collisions, the $n=25$ node is also driven.

For the sake of convenience towards the theoretical analysis, we further approximate the lattice as an $\alpha$-FPUT lattice by Taylor expanding Eq.~\eqref{eq:force_relation} up to the second order and non-dimensionalize using the following scaling.
\begin{align}
    u\rightarrow\dfrac{u}{D_0},\quad t\rightarrow t\omega_0,
\end{align}
where $\omega_0=\sqrt{\dfrac{a_2}{m}}$ and $a_2$ is the linear coefficient of the Taylor expansion~(details can be found in Supplementary Note~4).
Now, Eq.~\eqref{eq:eom_hinge} becomes,
\begin{align}\label{eq:fput_rescale}
    &\ddot{u}_n+\alpha_2\left(-u_{n+1}+2u_n-u_{n-1}\right)\nonumber\\
        &\quad+\alpha_3\left[(u_n-u_{n+1})^2-(u_{n-1}-u_n)^2\right]
        +2\zeta\dot{u}_n
        =0,
\end{align}
where $\alpha_{2}=1$, $\alpha_3=\dfrac{K_{3}D_0}{K_{2}}$, and $\zeta=\dfrac{c_{\rm damp}}{2\sqrt{K_{2}m}}$ with $K_{2}$ and $K_{3}$ being linear and quadratic coefficient of Taylor expansion.
(See Supplementary Note~4 for the explicit expression of $K_{2}$ and $K_{3}$.)

\subsubsection{Linear wave dispersion relationship}
We first inspect the agreement between the mathematical model and the experimental setup in the dynamic regime by examining the linear wave dispersion relationship.
Substituting the Bloch-Floquet ansatz into Eq.~\eqref{eq:fput_rescale} and solving for the angular frequency $\omega_{\rm L}$ yields
\begin{align}\label{eq:dispersion_linear}
    {\omega_{\rm L}}^2=4\alpha_2\sin^2\left(\frac{k}{2}\right),
\end{align}
where $k$ is the wavenumber.
In Fig.~\ref{fig:schematic}h, we show the dispersion relationships from the analytical prediction~[cf. Eq.~\eqref{eq:dispersion_linear}] and from the experimental result, denoted as the black dashed line and color intensity of the orange surface map, respectively.
The experimental result is obtained from a frequency sweep test~(excitation on both boundaries with $1$ to $101$ Hz for $40$ s), processed with a two-dimensional fast Fourier transform.
The infinite chain dispersion relationship matches well the high-intensity region of the experimental result, showing the validity of our mathematical model in the small amplitude regime for the lattice.
Please see Supplementary Note~5 and Supplementary Figure~5 for detailed experimental results of the frequency sweep test.

\subsection{Multiple-scale expansion and boussinesq approximation}\label{sec:approximation}

With a simplified model at hand and its validation in the linear limit, we now proceed to study the nonlinear dynamics of the system.
In particular, our focus will be on the interaction of rarefaction solitary waves.
Despite its simplified form, there are no analytical solutions of Eq.~\eqref{eq:fput_rescale} available, and so we must resort to approximation methods. 
While a reduction to the Korteweg-De Vries equation (KdV) equation using a multiple-scale expansion is a natural choice, the KdV equation is only unidirectional, and hence cannot capture collisions, which 
constitute the focus of this paper. 
We will still employ a multiple-scale expansion, but we aim to derive a bi-directional PDE, namely a Boussinesq equation. 
We detail that analysis here.

We assume the lattice constant is much smaller than the characteristic width of solitary waves, which we further discuss in detail in the next section.
Specifically, we  employ the following perturbation with a perturbation parameter $0<\epsilon\ll1$:
\begin{align}\label{eq:boussinesq_u}
    u=\epsilon A\left(\xi,\tau\right),
\end{align}
where
\begin{align}\label{eq:boussinesq_xitau}
    \xi=\epsilon n,\quad \tau=\epsilon t,
\end{align}
are long spatial and slow temporal variables, respectively.
Additionally, we scale the damping factor as,
\begin{align}\label{eq:boussinesq_gamma}
    \gamma=\frac{\zeta}{\epsilon},
\end{align}
to derive a long-wavelength damped-Boussinesq approximation. 
The displacement of the neighboring particles is approximated using the Taylor series,
\begin{align}
    u_{n\pm1}\left(t\right)
    \approx
    \sum_{i=0}^{4}\left(-1\right)^i\dfrac{\epsilon^i}{i!}\partial^iu.
\end{align}
Substituting Eqs.~\eqref{eq:boussinesq_u}, \eqref{eq:boussinesq_xitau}, and \eqref{eq:boussinesq_gamma} into Eq.~\eqref{eq:fput_rescale}, and collecting the terms at $\mathcal{O}\left(\epsilon^2\right)$ and $\mathcal{O}\left(\epsilon^4\right)$ yields the damped-Boussinesq equation,
\begin{align}\label{eq:boussinesq}
    \partial_\tau^2A+\nu_2\partial_\xi^2A+\nu_3\partial_\xi A\partial_\xi^2A+\nu_4\partial_\xi^4A+\nu_5\partial_\tau A=0,
\end{align}
or using the strain function $q(\xi,\tau)=-\partial_\xi A$,
\begin{align}\label{eq:boussinesq_strainform}
    \partial_\tau^2q+\nu_2\partial_\xi^2q-\frac{1}{2}\nu_3\partial_\xi^2\left(q^2\right)+\nu_4\partial_\xi^4q+\nu_5\partial_\tau q=0.
\end{align}
Here, the coefficients of the Boussinesq equation are related to the lattice parameters as follows:
\begin{align}
    \nu_2=-\alpha_2,\quad
    \nu_3=2\alpha_3\epsilon^2,\quad
    \nu_4=-\dfrac{1}{12}\alpha_2\epsilon^2,\quad
    \nu_5=\dfrac{2\zeta}{\epsilon}.
\end{align}
Note that using our strain-softening lattice parameters, we have that $\alpha_2>0$ and $\alpha_3<0$. In what follows, we focus on soliton solutions of Eq.~\eqref{eq:boussinesq_strainform}, rather than solutions resulting from the corresponding
initial value problem. 
This will help avoid issues stemming from the ill-posedness of Eq.~\eqref{eq:boussinesq_strainform}.
This is analogous to what has been done in earlier works working with ill-posed models, see \cite{Nesterenko2001} for a notable example.
Nonetheless, we will also consider numerical simulations of the lattice model to validate the derivation, as discussed below.

\subsection{One-soliton dynamics}\label{sec:onesoliton}

The Boussinesq equation is known to have exact soliton solutions in the undamped limit, the prototypical one of which can easily be obtained by setting $\zeta=0$~\cite{Ablowitz2011a} as
\begin{align}\label{eq:boussinesq_soliton_q}
    q=-a\sech^2\left[b\left(\xi-c\tau\right)\right],
\end{align}
where $a$ is the amplitude of the solitary wave, and the width and speed of the wave can be determined as,
\begin{subequations}\label{eq:soliton_parameter}
\begin{align}
    b&=\sqrt{\frac{a}{12}\frac{\nu_3}{\nu_4}}=\sqrt{-2a\frac{\alpha_3}{\alpha_2}},\label{eq:soliton_parameter_b}\\
    c&=\pm\sqrt{-\nu_2-\frac{a\nu_3}{3}}=\sqrt{\alpha_2-\frac{2}{3}a\alpha_3\epsilon^2}.\label{eq:soliton_parameter_c}
\end{align}
\end{subequations}
Therefore, the approximate solution for the lattice can be found as
\begin{align}\label{eq:boussinesq_soliton_du}
    \Delta u_n(t)=-\epsilon a\sech^2\left[\epsilon b\left(n-ct\right)\right],
\end{align}
or in terms of displacement,
\begin{align}\label{eq:boussinesq_soliton_u}
    u_n\left(t\right)=\epsilon\frac{a}{b}\tanh\left[\epsilon b\left(n-ct\right)\right].
\end{align}
Notice in Eqs.~\eqref{eq:boussinesq_soliton_du} and \eqref{eq:boussinesq_soliton_u} that the small parameter $\epsilon$ and soliton parameter $a$ always appear together as the product $\epsilon\sqrt {a}$.
Thus, we will set $\epsilon=1$ and consider $a$ the underlying parameter.
Here, the pulse width of the soliton solution in 
the spatial domain can be determined as $\Lambda=\dfrac{2\cosh^{-1}{\sqrt{2}}}{\epsilon b}$ which corresponds to the full width at half maximum~(FWHM), a standard diagnostic for characterizing the 
spatial extent of solitary waves~\cite{Li2018c, Sciacca2020}.
Note that for our lattice, the FWHM can be estimated to be $\Lambda\approx3.2$~(equivalent to approximately $96$ mm in dimensional units) when $a=0.1$ and $\epsilon=1$.
While an FWHM of $96$ mm is larger than the lattice constant $D_0=30$~mm of our lattice, the two are of the same order of magnitude.
It is often the case that multiple-scale approximations are accurate beyond the assumptions under which they were derived~\cite{Wright2014, Ari2024, NLWbook2024}.
Nonetheless, the validity of the multiple-scale approximation employed here must be corroborated against numerical simulations, which will be discussed next.

\begin{figure*}[t]
    \centering
    \includegraphics[width=\linewidth]{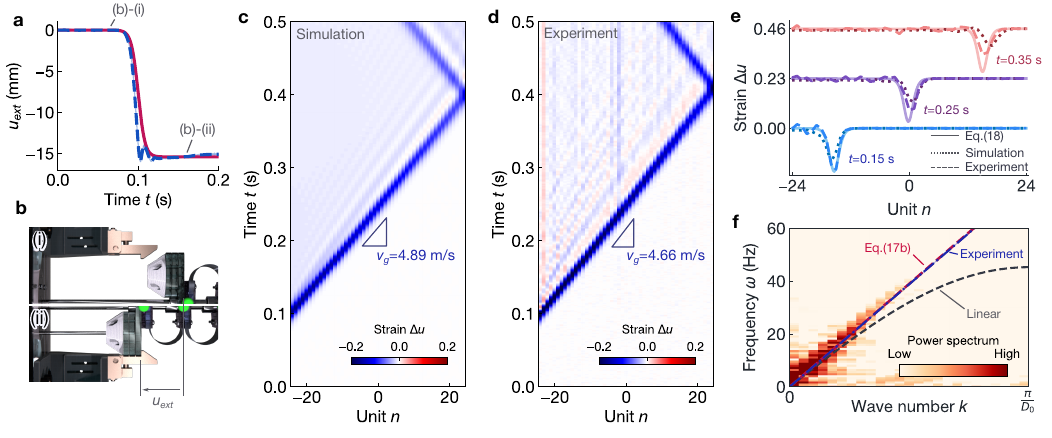}
    \caption{Single hump solitary wave solutions.
    (\textbf{a}) Excitation profile based on the analytical solution~[cf. Eq.~\eqref{eq:boussinesq_soliton_u}].
    Red solid line, analytical solution;
    blue dashed line with a shaded area represents the average and standard deviation of excitation in the experiment.
    (\textbf{b}) Schematical illustration of $u_{ext}$  for $t<0.1$ s (sub panel (i)) and $t>0.1$ s (sub panel (ii)) overlaid on the top view of boundary structure shown in Fig.~\ref{fig:schematic}b.
    (\textbf{c}) Damped full lattice simulation~[cf. Eq.~\eqref{eq:eom_hinge}].
    (\textbf{d}) Experimental result.
    The amplitude of the soliton input is set to $a=0.2$.
    (\textbf{e}) Comparison of the spatial profile at $t=0.15$, $0.25$, and $0.35$ s.
    Solid lines, analytical solution;
    dotted lines, lattice simulation;
    dashed lines, experiment.
    The color scheme goes as blue, purple, and red for $t=0.15$, $0.25$, and $0.35$ s, respectively.
    The y-axis is shifted by 0.23 each for better visibility.
    (\textbf{f}) Dispersion relationships.
    Black dashed line, linear wave dispersion relationship~[cf. Eq.~\eqref{eq:dispersion_linear}];
    red dashed line, group velocity given by Eq.~\eqref{eq:soliton_parameter_c};
    blue dash-dotted line, solitary wave speed estimated from the experiment;
    color intensity, 2D FFT spectrum of experimental result.
    }
    \label{fig:onesoliton}
\end{figure*}

In Fig.~\ref{fig:onesoliton}, we compare the analytical solution~[cf. Eq.~\eqref{eq:boussinesq_soliton_q}], full nonlinear lattice simulation~[cf. Eq.~\eqref{eq:eom_hinge}], and the experimental results of the lattice~[cf. Fig.~\ref{fig:schematic}a].
See Supplementary Table~1 for the parameter values used.
In the lattice simulation and experiment, the left boundary of the lattice undergoes tensile excitation to generate a solitary wave propagating to the right.
In the simulation, this is achieved by simply setting $u_0(t)=u_{\rm ext}$ in Eq.~\eqref{eq:eom_hinge}, where $u_{\rm ext} = \epsilon\frac{a}{b}\tanh\left[-\epsilon bct\right]$ is the analytical rarefaction wave with $n=0$. For the experiment, we fed the same excitation to a function generator using a MATLAB script, which was then sent to the shaker.
See Fig.~\ref{fig:onesoliton}a for a comparison of the initial excitation in simulation and experiment.
In the experiment, after the boundary particle is displaced to the left and reaches a predefined displacement, a pair of latches holds the particle in place, as shown in Fig.~\ref{fig:onesoliton}b.
This prevents any ``backlash'' compressive waves from being induced, leaving only tensile waves within the lattice~(see Supplementary Video~1 for the details of the latch operation).

In Fig.~\ref{fig:onesoliton}c and \ref{fig:onesoliton}d, we show the spatiotemporal landscape of the strain wave field from the numerical simulation and experiment, respectively~(for lattice motion in the experiment, see Supplementary Video~4).
For both simulation and experiment, the rarefaction waves are triggered at $0.1$ s and reach the right boundary at approximately $0.4$ s.
An oscillatory tail follows the rarefaction waves.
In the simulation, this is due to the deviation from the approximate solution given by the Boussinesq equation.
As $\epsilon\rightarrow0$, the tails become less evident~(see Supplementary Note~6 and Supplementary Figure~6 for the numerical solutions with different values of $\epsilon$).
For the experimental result, two other factors play a critical role in creating trailing waves.
As Fig.~\ref{fig:onesoliton}a shows, the input signal contains small oscillatory noise around $u_{\rm ext}=-15$, which is caused by the latch motion, generating additional small impulses from the boundary.
Additionally, the effect of the damping plays a role, leaving a uniform, small-amplitude tension remaining after the rarefaction wave passes through the lattice instead of each of the particles fully returning to the original position~(see Supplementary Note~6 and Supplementary Figure~6 for the comparison between damped and undamped lattice simulation).

Figure~\ref{fig:onesoliton}e compares the spatial profiles of solitary waves in detail.
We can see that the lattice solitary wave profiles~(simulation and experiment) match well with the Boussinesq simulation in both amplitude and width.
(For experimental results for different amplitudes, $a=0.05$, $0.1$, $0.15$, and $0.25$, see Supplementary Note~7 and Supplementary Figures~7 and 8.)

We present the dispersion relationship in Fig.~\ref{fig:onesoliton}f to further confirm that the experimental strain profile is a solitary wave.
The color intensity is concentrated at the zero wave number, with a slope slightly greater than the group velocity, as expected for super-sonic solitary waves.
The group velocity determined from linear dispersion relatinship~[Eq.~\eqref{eq:dispersion_linear}] is $4.37$ m/s, whereas the speed of the solitary wave is $4.66$ m/s for experiment~(average of 10 trials with standard deviation of $8.32\times10^{-3}$ m/s), $4.68$ m/s based on Eq.~\eqref{eq:soliton_parameter_c}, and $4.89$ m/s in the simulation.
This is in contrast to the linear wave dispersion relationship described by Eq.~\eqref{eq:dispersion_linear} and in Fig.~\ref{fig:schematic}f, denoted as a black dashed line.
(See Supplementary Table~2 for agreement of solitary wave speed between analytical and experimental results.)

\subsection{Head-on collision of rarefaction solitons}\label{sec:twosoliton}

The solitary wave collision experimental result can be obtained by following the same procedure as the single solitary wave experiment, but now with the shaker driving both left and right boundaries.
Here, we consider two cases: (i) symmetric collision with $a_1=0.2$, $a_2=0.2$, and (ii) asymmetric cases with $a_1=0.1$, $a_2=0.2$.

\begin{figure*}[tbp]
    \centering
    \includegraphics[width=\linewidth]{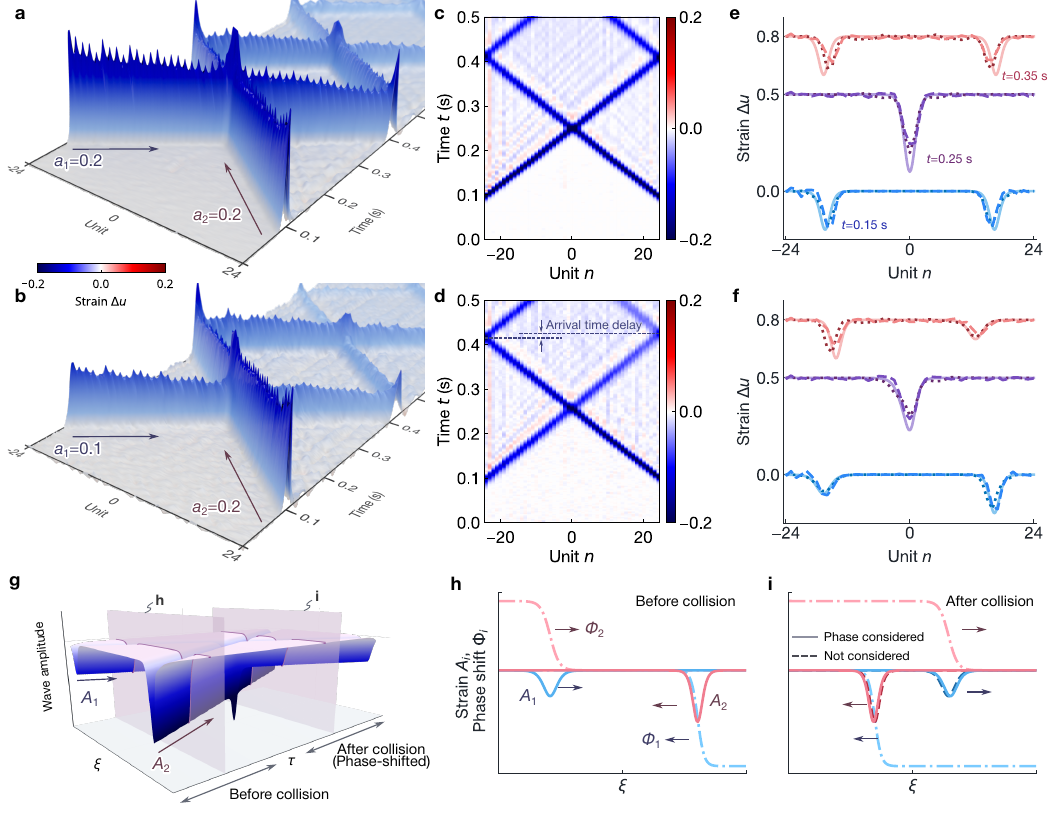}
    \caption{Head-on-collision of two rarefaction solitary waves in the experiment.
    Three-dimensional view visualizing (\textbf{a}) a symmetric collision with $a_1=a_2=0.2$, and (\textbf{b}) an asymmetric collision with $a_1=0.1$, $a_2=0.2$.
    Two-dimensional surface plot visualizing spatiotemporal trajectory for the (\textbf{c}) symmetric and (\textbf{d}) asymmetric collision case.
    Spatial profiles of the strain for the (\textbf{e}) symmetric and (\textbf{f}) asymmetric cases.
    The color scheme goes as blue, purple, and red for $t=0.15$, $0.25$, and $0.35$ s, respectively.
    Dashed lines: experiment; dotted lines: damped simulation; solid lines: theory.
    Schematic illustration of the phase shift shown in panels (g)-(i).
    (\textbf{g}) Spatiotemporal landscape of wave amplitude $A_1+A_2$ under head-collision.
    Wave amplitudes $A_i$ and phase shifts $\Phi_i$ 
    (\textbf{h}) before and 
    (\textbf{i}) after the collision.
    Solid lines, wave amplitude $A_i$;
    dashdotted lines, phase shift $\Phi_i$;
    dashed lines, wave amplitude $A_i$ without phase shift taken into account~[Eq.~\eqref{eq:boussinesq_soliton_q}].
    Blue lines, right-going soliton;
    red lines, left-going soliton.
    }
    \label{fig:twosoliton}
\end{figure*}

Panels~\ref{fig:twosoliton}a and \ref{fig:twosoliton}b show the spatiotemporal landscape of the strain wave field for symmetric and asymmetric collision cases, respectively~(for the lattice motion during the experiment, see Supplementary Video~5).
We can clearly see the formation of large amplitude waves upon collision near $n=0$ for both cases.
Furthermore, the collision is repeated multiple times after the solitary waves are reflected at the boundary while slowly getting attenuated due to the damping.
In regard to the difference between symmetry and asymmetry, a subtle contrast between them can be observed by closely examining Fig.~\ref{fig:twosoliton}c and \ref{fig:twosoliton}d.
For the symmetric case shown in Fig.~\ref{fig:twosoliton}c, two solitary waves collide at $n=0$ and reach the opposite boundary at the same moment around $t=0.42$ s.
On the contrary, Fig.~\ref{fig:twosoliton}d shows the right-going solitary wave reaches the other end slightly later than the left-going solitary wave, suggesting the difference in speed, which stems from the difference in excitation amplitude as indicated by two horizontal dashed lines.
Specifically, travel times for right- and left-going solitary waves to reach the end of the lattice are $0.326$ s and $0.310$ s on average~(based on 10 trials with a standard deviation of $1.70\times10^{-3}$ and $1.01\times10^{-3}$ s), respectively.
(Experimental results with other amplitudes of left-going solitary waves can be found in Supplementary Figures~7 and 8.)

Figures~\ref{fig:twosoliton}e and \ref{fig:twosoliton}f compare the spatial profile of the strain wave between experiment, simulation, and theory.
Specifically, we extract the strain profiles before, at, and after the collision.
The simulation results are obtained by numerically solving Eq.~\eqref{eq:eom_hinge} with damping, as before, but now the right boundary $u_{25}(t)$ is given by Eq.~\eqref{eq:boussinesq_soliton_u} with $n=25$.
The analytical curves are derived through multiple-scale analysis, but unlike the one-soliton case, we now need to account for the soliton interaction.
In other dispersive wave settings (e.g., in optics~\cite{Kivshar2003}), solitary wave collisions are nearly elastic (when close to the integrable regime), but each wave becomes phase-shifted~\cite{Ablowitz2011a}.
Thus, we will return to our multiple-scale analysis but include terms that will account explicitly for the phase shift. A similar strategy has been employed in optical settings~\cite{CBSU2007} and in lattice settings~\cite{SUW2011}. 

To derive the two-soliton solution with phase shift, we employ a traveling wave ansatz with two wave speeds $c_1$ and $c_2$, such that
\begin{align}\label{eq:disp_twosoliton}
    u_n&=\epsilon A_1(X_1)+\epsilon A_2(X_2),
\end{align}
where
\begin{subequations}
\begin{align}
    X_1=\xi-c_1\tau+\epsilon^2\Phi_1(\widetilde{X}_2),&\quad
    X_2=\xi-c_2\tau+\epsilon^2\Phi_2(\widetilde{X}_1),\\
    \widetilde{X}_1=\xi-c_1\tau,&\quad
    \widetilde{X}_2=\xi-c_2\tau.
\end{align}    
\end{subequations}
As before $\xi = \epsilon n$ and $\tau = \epsilon t$.
Here, $A_1$ will represent the right-moving wave, and $A_2$ will represent the
left-moving wave~(see Fig.~\ref{fig:twosoliton}g for the expected spatiotemporal landscape with two waves).
The phase correction term for $A_1$ is $\Phi_1(\tilde{X}_2)$.
Note that the phase correction depends on a traveling coordinate with speed corresponding to $A_2$, namely $c_2$.
When the waves are well separated, we want the phase correction to be near zero.
Once the waves interact, we want the phase shift to come into play, hence the dependence
of $\Phi_1$ on $c_2$.
The same reasoning is applied to the phase
correction $\Phi_2(\tilde{X}_1)$, see Fig.~\ref{fig:twosoliton}h and \ref{fig:twosoliton}i for an illustration.
Note that $X_1=\widetilde{X}_1+\mathcal{O}(\epsilon^2)$.
For the sake of derivation simplicity, we use the approximation $ X_1 = \widetilde{X}_1$, and likewise for $X_2$, which will be shown to have a fair accuracy later in the section.
Subsituting the ansatz Eq.\eqref{eq:disp_twosoliton} into Eq.~\eqref{eq:fput_rescale} yields the following
\begin{widetext}
\begin{align}\label{eq:twosoliton_ode}
    {c_1}^2{A_1}^{\prime\prime}\left(X_1\right)-\alpha_2{A_1}^{\prime\prime}\left(X_1\right)+2\alpha_3\epsilon^2{A_1}^{\prime}\left(X_1\right){A_1}^{\prime\prime}\left(X_1\right)-\frac{1}{12}\alpha_2\epsilon^2{A_1}^{(4)}\left(X_1\right)&\nonumber\\
    +{c_2}^2{A_2}^{\prime\prime}\left(X_2\right)-\alpha_2{A_2}^{\prime\prime}\left(X_2\right)+2\alpha_3\epsilon^2{A_2}^{\prime}\left(X_2\right){A_2}^{\prime\prime}\left(X_2\right)-\frac{1}{12}\alpha_2\epsilon^2{A_2}^{(4)}\left(X_2\right)&\nonumber\\
    + \epsilon^2\left[
    -2\alpha_3{B_1}^{\prime}(X_2){A_1}^{\prime\prime}(X_1)
    +2\left(\alpha_2-c_1c_2\right){\Phi_1}^{\prime}(X_2){A_1}^{\prime\prime}(X_1)
    +\left(\alpha_2-{c_2}^2\right){\Phi_1}^{\prime\prime}(X_2){A_1}^{\prime}(X_1)
    \right]&\nonumber\\
    + \epsilon^2\left[
    -2\alpha_3 {A_1}^{\prime}(X_1){B_1}^{\prime\prime}(X_2)
    +2\left(\alpha_2-c_1c_2\right){\Phi_2}^{\prime}(X_1){B_1}^{\prime\prime}(X_2)
    +\left(\alpha_2-{c_1}^2\right){\Phi_2}^{\prime\prime}(X_1){B_1}^{\prime}(X_2)
    \right]&=0
\end{align}
\end{widetext}
Note that the first two terms are linear combinations of the Boussinesq approximation~[cf. Eq.~\eqref{eq:boussinesq}]. 
The last two terms result from the interaction of the two waves (namely, their coupling via the nonlinearity of the system).
To eliminate the last two terms, we assume the following relationships:
\begin{subequations}\label{eq:phi2amplitude}
\begin{align}
    \Phi_1(X_2)=\gamma_1A_2(X_2)+{\Phi_1}^{(0)}\\
    \Phi_2(X_1)=\gamma_2A_1(X_1)+{\Phi_2}^{(0)}
\end{align}
\end{subequations}
where $\gamma_1$,$\gamma_2$ are constants to be determined, and $\Phi_1^0$ and $\Phi_2^0$ are the initial phases.

If we let $A_1$ and $A_2$ be the solitary wave solutions 
[see Eq.~\eqref{eq:boussinesq_soliton_u}] and assume the relations in Eq.~\eqref{eq:phi2amplitude}, then Eq.~\eqref{eq:twosoliton_ode} reduces to,
\begin{align}
\epsilon^2{A_2}^{\prime}{A_1}^{\prime\prime}
\left(2\alpha_2\gamma_1+\alpha_2\gamma_2-2\alpha_3-2c_1c_2\gamma_1-{c_1}^2\gamma_2\right)&\nonumber\\
+\epsilon^2{A_1}^{\prime}{A_2}^{\prime\prime}
\left(2\alpha_2\gamma_2+\alpha_2\gamma_1-2\alpha_3-2c_1c_2\gamma_2-c_2^2\gamma_1\right)&=0.
\end{align}
This leads to the linear system
\begin{subequations}\label{eq:phase_correction}
\begin{align}
    2(c_1c_2-\alpha_2)\gamma_1+(c_1^2-\alpha_2)\gamma_2&=-2\alpha_3\\
    2(c_1c_2-\alpha_2)\gamma_2+(c_2^2-\alpha_2)\gamma_1&=-2\alpha_3,
\end{align}
\end{subequations}
which has the solution
\begin{subequations}
\begin{align}
      \gamma_1 &=
\frac{2 \left(\alpha_2  \alpha_3 +\alpha_3  c_1^2-2 \alpha_3 
   c_1 c_2\right)}{3 \alpha_2 ^2+\alpha_2 
   c_1^2-8 \alpha_2  c_1 c_2+\alpha_2 
   c_2^2+3 c_1^2 c_2^2} \\
  \gamma_2 &=
   \frac{2 \left(\alpha_2  \alpha_3 -2 \alpha_3  c_1
   c_2+\alpha_3  c_2^2\right)}{3 \alpha_2
   ^2+\alpha_2  c_1^2-8 \alpha_2  c_1
   c_2+\alpha_2  c_2^2+3 c_1^2
   c_2^2}.
\end{align}
\end{subequations}

There should be no phase shift before the head-on collision, and thus, we impose
\begin{subequations}\label{eq:phi}
\begin{align}
\left.\Phi_1(X_2)\right|_{X_2\rightarrow-\infty}&=
    \left.\gamma_1A_2(X_2)\right|_{X_2\rightarrow-\infty}+{\Phi_1}^{(0)}=0,\\
\left.\Phi_2(X_1)\right|_{X_1\rightarrow\infty}&=
    \left.\gamma_2A_1(X_1)\right|_{X_1\rightarrow\infty}+{\Phi_2}^{(0)}=0,
\end{align}
\end{subequations}
which implies,
\begin{align}\label{eq:phi0}
    {\Phi_1}^{(0)}=\gamma_1\frac{a_2}{b_2},\quad
    {\Phi_2}^{(0)}=\gamma_2\frac{a_1}{b_1}
\end{align}
Therefore, an estimate for the phase shift after the interaction can be determined as
\begin{align}\label{eq:phaseshift}
    \max\Phi_1=2\epsilon\gamma_1\frac{a_2}{b_2},\quad
    \max\Phi_2=2\epsilon\gamma_2\frac{a_1}{b_1}.
\end{align}

The analytical solution with the phase shift function satisfying Eq.~\eqref{eq:phi} substituted into Eq.~\eqref{eq:disp_twosoliton} is used to obtain the strain profile as shown in Figs.~\ref{fig:twosoliton}e and (f), denoted as solid lines.
We can see that the experiment and simulation results coincide with the analytical prediction. 

\begin{figure*}[t]
    \centering
    \includegraphics[width=\linewidth]{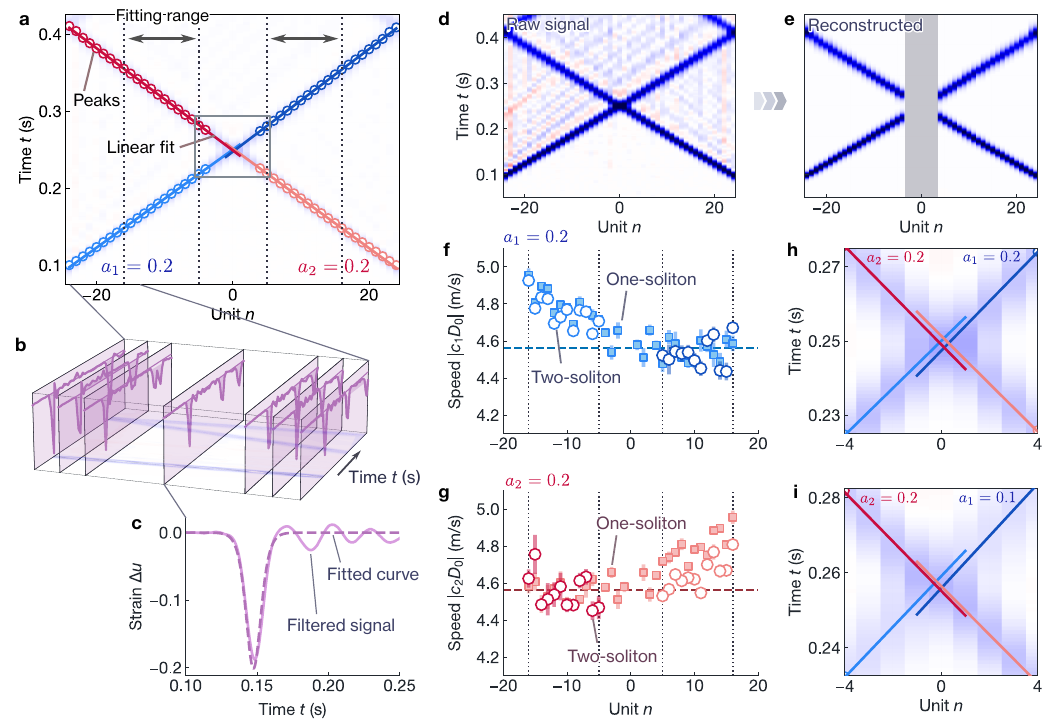}
    \caption{Phase shift due to the head-on collision.
    (\textbf{a}) Trajectories of the solitary waves.
    Open circle symbols: peak location of solitary waves;
    Solid lines: trajectories estimated by linearly fitting peaks.
    Blue: right-going solitary wave;
    Red: left-going solitary wave.
    Light and dark colors represent peaks and trajectories before and after the collision.
    (\textbf{b}) Slices of time series data at each unit cell upon fitting of Eq.~\eqref{eq:boussinesq_soliton_q} to the experimental result.
    (\textbf{c}) Temporal profile of experimental data~(solid line) with a low-pass filter and fitted curve~(dashed line) at $n=-17$.
    Surface map of (\textbf{d}) raw strain wave data and (\textbf{e}) reconstructed data.
    Speed of (\textbf{f}) right-going and (\textbf{g}) left-going solitary waves compared to the single solitary wave experimental result and the approximate solution.
    Dashed lines: speed based on Eq.~\eqref{eq:soliton_parameter_c};
    Open circle symbols: collision experiment;
    Filled square symbols: single solitary wave experiment.
    Error bars represent the standard deviation.
    The light and dark colors of the circle symbols correspond to solitary waves before and after the collision, respectively, following the color convention in panel (\textbf{a}).
    Trajectories of the solitary waves near $n=0$ for
    (\textbf{h}) symmetric~($a_1=a_2=0.2$) and (\textbf{i}) asymmetric case~($a_1=0.1,\,a_2=0.2$).
    The solid lines correspond to a linear fit of the trajectories.
    }
    \label{fig:fitting}
\end{figure*}

Now, we further assess the experimental result by examining the properties of the (near) elastic collision, namely the speed and phase shift.
To this end, we first extract the trajectory of each solitary wave before and after the collision.
In Fig.~\ref{fig:fitting}a, we show the trajectory of symmetrically colliding solitary waves by marking the location of the peak of the solitary waves for each unit $n$ as open circle symbols.
The peak location is determined through the following process:
we extract time series data for each unit cell, as illustrated in Fig.~\ref{fig:fitting}b.
Each set of temporal data is individually processed with a low-pass filter, then used for fitting the one-soliton equation, slightly modified for convenience from Eq.~\eqref{eq:boussinesq_soliton_q} as
\begin{align} \label{eq:fit}
    f_{\rm fit}\left(t;p_1,p_2,t_0\right)=p_1\sech^2\left[p_2\left(t-t_0\right)\right],
\end{align}
where $p_1$, $p_2$, and $t_0$ are fitting parameters that correspond to soliton amplitude $a$, product of width and speed $bc$, and temporal shift.
The result of the fitting is shown in Fig.~\ref{fig:fitting}c.
The fitted curve effectively isolates a solitary wave of the $\sech$ type while filtering out the oscillatory tail.
The spatiotemporal field can be reconstructed by mapping the result as shown in Figs.~\ref{fig:fitting}d and \ref{fig:fitting}e, which displays only two solitary waves without background noise.
The area around the collision site, indicated by the grey-shaded area, is intentionally excluded from the curve fitting to enhance accuracy, since collisions can distort the fitting process.
Fitting the data in this way not only filters out noise but also helps us determine the peak location accurately.
The true peak may fall between where the raw data is sampled (i.e., at multiples of the frame duration).
Fitting the data with a trial function allows the peak location to take on a continuum of values, which would not be possible if finding the maximum of the sampled data. 
For the trial function given in Eq.~\eqref{eq:fit}, $t_0$ represents the position of the peak. 
In Fig.~\ref{fig:fitting}a, the $t$ coordinate of the open circles corresponds to the best-fit value of $t_0$ for each unit $n$ in the fitting area.
(For more details of the fitting results, see Supplementary Note~8 and Supplementary Figure~9.)

As can be inferred from the spatiotemporal landscape in Figs.~\ref{fig:fitting}d and \ref{fig:fitting}e, the trajectory represented by a series of peak locations involves a practically linear (constant speed) profile before and after the collision for both solitary waves.
Here, the region near the collision is not considered for identifying the trajectory, as the purpose is to extract pre- and post-collision trajectories separately to be compared.

Based on the extracted trajectories, the speed of the solitary waves can be estimated, as shown in Fig.~\ref{fig:fitting}f and \ref{fig:fitting}g, denoted as open circle symbols~(results are average of 10 trials, the standard deviation of which are indicated as error bars).
The speed of the solitary waves remains approximately constant throughout the trial, except for the slow decay due to the energy dissipation, which occurs regardless of the collision~(i.e., the speed slightly decreases even before the collision).
Furthermore, the decrease in the speed for the head-on collision case follows the same trend as that of the single solitary wave case denoted as filled square symbols in Fig.~\ref{fig:fitting}f and \ref{fig:fitting}g~(approximately $4.46$\% and $4.78$\% of decrease from the leftmost to the rightmost markers, respectively).
This suggests that the collision is essentially elastic, and the decrease in speed is not significantly affected by the collision.
(See Supplementary Table~3 for agreement of solitary wave speed between analytical and experimental results.)

To quantify the shift in space and time, we conduct linear fitting for each trajectory, which is denoted as solid lines overlaid on the trajectory in Fig.~\ref{fig:fitting}a.
Note that we fit within the region between the black dotted lines indicated in Fig.~\ref{fig:fitting}a to exclude the trajectory in the vicinity of $n=\pm24$ and $n=0$, such that the linear fitting shall avoid the effect of the boundary and collision.
Figure~\ref{fig:fitting}h shows the magnified view around the collision point~[i.e., the region enclosed by a grey rectangle in Fig.~\ref{fig:fitting}a].
We can clearly see lines being shifted after going through the collision, while the slope of the linear fitting remains approximately the same~[which matches with the observation from Fig.~\ref{fig:fitting}f and \ref{fig:fitting}g].
Furthermore, as the symmetric collision suggests, the amount of the shift is approximately equal for two solitary waves.
This is not the case for the asymmetric collision, which can be confirmed from Fig.~\ref{fig:fitting}i, where the asymmetric collision between $a_1=0.1$ and $a_2=0.2$ is shown.
Here, we see an asymmetry in the shift of the trajectory.
In particular, the right-going solitary wave shifts more than the left-going (taller) solitary wave.

\begin{figure*}[t]
    \centering
    \includegraphics[width=\linewidth]{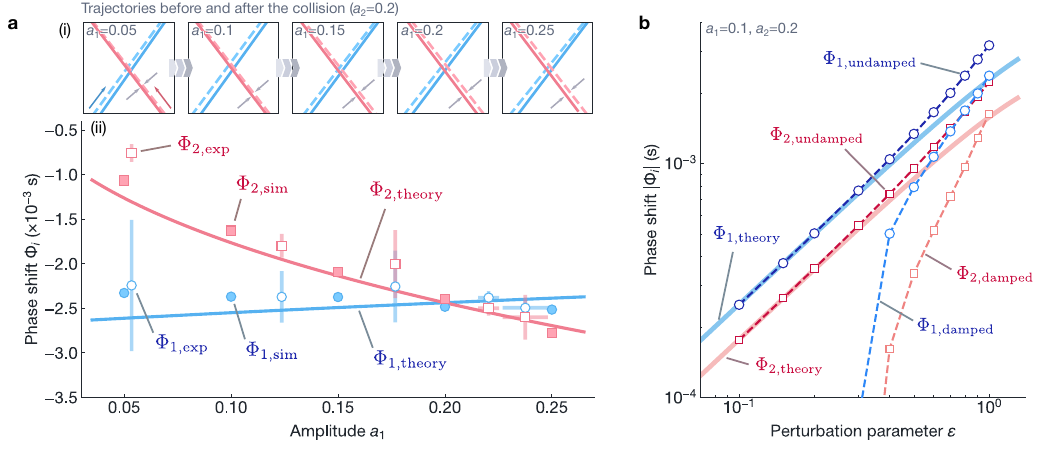}
    \caption{Phase shift due to the head-on collision.
    (\textbf{a}) Phase shift as a function of the right-going solitary wave amplitude $a_1$.
    (i) Trajectories before (dashed) and after (solid) the collision for different $a_1$.
    (ii) Comparison between theory, numerical simulation, and experiment.
    Open symbols: experimental results;
    Filled symbols: damped simulation results with Eq.~\eqref{eq:eom_hinge};
    Solid lines: the analytical solution of Eq.~\eqref{eq:phaseshift}.
    Error bars represent standard deviation.
    (\textbf{b}). Phase shift $\left|\Phi_1\right|$ as a function of the perturbation parameter $\epsilon$.
    The amplitudes of the right- and left-going solitary waves are kept constant at $a_1=0.1$ and $a_2=0.2$.
    }
    \label{fig:phaseshift}
\end{figure*}

Figure~\ref{fig:phaseshift}a further examines the phase shift as a function of the amplitude of the soliton.
The head-on collision experiment is repeated by varying the amplitude of the right-going solitary waves as $a_1=0.05,\,0.1,\,0.15,\,0.2$, and $0.25$ while keeping the left-going solitary wave at $a_2=0.2$.
In subpanel (i), we present trajectories of solitary waves for different amplitudes in the experiment~[panels with $a_1=0.1$ and $a_1=0.2$ are reprinted from Fig.~\ref{fig:fitting}h and (i)].
If we examine the qualitative behavior of the solitary waves, we can see that right-going solitary waves show a similar gap distance between the trajectories before~(dashed) and after~(solid) the collision.
On the contrary, the left-going solitary waves represented by the red lines show an increase in the gap between the trajectories before and after the collision.
Such difference in the qualitative behavior can be confirmed in Fig.~\ref{fig:phaseshift}a, sub-panel (ii).
The experimental result, marked with open symbols, shows the decrease of $\Phi_2$ as $a_1$ increases, as opposed to the almost constant $\Phi_1$ regardless of $a_1$.
This matches the trend of the numerical results and analytical prediction~[cf. Eq.~\eqref{eq:phaseshift}], denoted as filled symbols and solid lines, respectively. 
The validity of the analytical prediction is further assessed by comparing the damped simulation, the undamped simulation, and Eq.~\eqref{eq:phaseshift}. 
In particular, we vary the speed while keeping the ratio $c_1/c_2$ constant by fixing $a_1=0.1$ and $a_2=0.2$ and varying the parameter $\epsilon$, see Fig.~\ref{fig:phaseshift}b.
We can see that the phase shifts $\Phi_1$ and $\Phi_2$ of the undamped simulation asymptotically approach the theoretical prediction as the perturbation parameter $\epsilon$ decreases, validating the multiple-scale approach for the interaction of two solitary waves.
Notice that the phase shift equation~[cf. Eq.~\eqref{eq:phaseshift}] describes the lattice behavior well, even for $\epsilon$ close to $1$.
On the other hand, the damped simulation is downshifted in comparison with the undamped simulation, resulting in the coincidence with the analytical solution closer to $\epsilon=1$.

\section{Discussion}\label{sec:discussion}
We have experimentally, numerically, and analytically demonstrated rarefaction wave interaction in an
unprecedented experimental strain-softening
platform and have calculated the associated phase shift in the mechanical lattice.
We first derived an effective Boussinesq equation to approximate the full lattice dynamics at its continuum limit.
To this end, we started by approximating our full lattice equation of motion in the form of an FPUT lattice, which we then converted to a Boussinesq equation through multiple-scale analysis.
The approximation based on the one-soliton solution of the Boussinesq equation compared favorably to both experimental and numerical results, except for the slight amplitude attenuation feature due to the energy dissipation (which was not included in the theoretical
analysis).
In the head-on collision case, the experimental and numerical results are validated with the analytical solution derived, where the interactions are considered by modifying the one-soliton approximation ansatz.
This is an important development enabling an analytical characterization of the emergent as measurable phase shifts of the soliton positions.
The resultant analytical solution describes the elastic collision of two solitons and the phase shift, which matches well both experimental and numerical results from both qualitative and quantitative aspects.
The experimental results are acquired through multiple data processing steps, including noise filtering, analytical solution fitting, and linear fitting, which successfully capture the millisecond order of phase shift.

We believe that these findings not only offer new insights into the dynamics of the soliton interactions and associated collision properties from both analytical and experimental perspectives but also provide a set of approaches that shall be effective in measuring, processing, and detecting such subtle physical phenomena within mechanical metamaterials. 
Naturally, numerous further perspectives emerge from this study. 
Among others, we note:
(1) further systematics of the role of dissipation
(which has been a topic of extensive interest; see, e.g.,
the recent analysis of~\cite{James_2021} and 
earlier works such as~\cite{Carretero-Gonzalez2009,vergara,lindenberg});
(2) the exploration of different (especially including slow) speeds;
(3) the potential for collisions off of regular lattice sites 
(similar to what was done, e.g., in~\cite{ioanna} that considered
intersite or even quarter-site collisions);
or (4) the more exotic scenarios of higher-soliton-collisions~\cite{dmitriev}, which can lead to intriguing radiationless energy exchange.
Such studies are currently under consideration and will be reported in future publications.


\section*{Methods}
\subsection{Fabrication and assembly}
Our system consists of lattice and modal shakers as discussed in the main text and Fig.~\ref{fig:schematic}.
The lattice consists of 49 unit cells~(i.e., 50 particles, including the boundary particles).
The unit cell consists of rectangular sheets of spring steel~(0.1 mm thick SK5 spring steel) connected through separators embedded with the ball-bearing~(Misumi U-LHFC0.25 Flanged Linear Ball Bearing).
Spring steel components, acting as nonlinear springs, are cut with the fiber laser cutter~(Bodor i6 Fiber Laser).
The ball-bearing housings are made of acrylic plates of two thicknesses~(4.5~mm and 1.6~mm), also processed with the CO$_2$ laser cutter~(Trotec Speedy 360).
Motion tracking markers are made of polytetrafluoroethylene~(PTFE) spheres with a $10$ mm diameter, colored with fluorescent green lacquer spray.
A pair of high-speed cameras~(Kron Technologies Chronos 1.4) track these markers at 10000 frames per second.
Therefore, the duration between the frames is $10^{-4}$ s, which is an order of magnitude smaller than the scale of phase shifts.     

Once assembled, the particles are inserted through a stainless steel shaft, which is supported with aluminum frames on the boundaries.
The boundary structure comes with a pair of latches that hold the boundary particle in place once the particle is displaced according to the solitary wave excitation profile~[see Fig.~\ref{fig:schematic}b in the main text and Supplementary Figure~1(c) sub-panel (i) for more detailed composition].
As discussed in the main text, to facilitate the latch operation, the particles on the boundaries~(i.e., $n=-24$ and $n=25$) are equipped with a wide acrylic plate~($5$ mm-thick), the width of which is slightly shorter than the distance between two latches~[see Fig.~\ref{fig:schematic}b, Fig.~\ref{fig:onesoliton}b, and Supplementary Figure~1(c) sub-panel (ii)].
The acrylic plate also serves as a connection between the lattice and the stinger via a 3D-printed stinger adaptor.
The side surface of the adaptor is polished multiple times with sandpaper of different grit sizes and treated with PTFE tape.
This prevents any unwanted friction between the latch claw and the stinger adaptor.

The stinger is inserted into the modal shaker~(Br\"uek\&Kj\ae r Modal Exciter 4825), which is connected to the function generator through the amplifier.
Note that for the symmetric collision case, two shakers are controlled by one function generator, while shakers are controlled by individual function generators in the asymmetric case.
For more details of geometry and assembly, see Supplementary Figure~1.

\subsection{Numerical methods}
\subsubsection{Numerical simulation}
As briefly mentioned in the main text, the lattice dynamic simulation is conducted by solving Eq.~\eqref{eq:eom_hinge} using the eighth-order Dormand-Prince method, with error estimation of a fifth and third order.
The maximum time step size is $\Delta t = 10^ {-6}$, and both absolute and relative tolerances are set to $10^{-12}$ to determine the time step.

The Boussinesq PDE, Eq.~\eqref{eq:boussinesq}, is also solved using the explicit method~(the results of which are mainly discussed in the Supplementary Note).
A sixth-order central difference scheme is employed for space discretization, and time integration is performed using the eighth-order Dormand-Prince method.
Here, the maximum time step size is set to $\Delta\tau=10^{-5}$ with absolute and relative tolerances of $10^{-12}$.

\subsubsection{Fitting}
The soliton solution fitting is performed using least-squares optimization with a Cauchy loss function to minimize the effect of outliers.
For robust fitting, 10\% of the total number of unit cells, specifically those centred around the interaction point, were excluded from this procedure.

A least-squares 1st order polynomial fitting was applied to extract the trajectories of solitary waves.
Similar to the siliton solution fitting, approximately 10\% of the unit cells near the collision point and on each boundary are excluded from the fitting process.

\section*{Data availability}
Data supporting the findings of this study are available in the Zenodo repository under the accession code https://doi.org/10.5281/zenodo.15301154.

\bibliography{main}

\begingroup
\setlength{\parindent}{0cm} 

\section*{Acknowledgments}
Y.M. and J.Y. acknowledge the support from SNU-IAMD,
SNU-IOER, and National Research Foundation grants funded by
the Korea government [Grants No. 2023R1A2C2003705 and No. 2022H1D3A2A03096579
(Brain Pool Plus by the Ministry of Science and ICT)].
We also thank Dr. Yeon June Kang and Mr. Jeong-min Nam at Seoul National University for kindly allowing us to use the modal shaker.
This material is based upon work supported by the U.S. National Science Foundation under the awards DMS-2107945 (C.C.) and PHY-2110030, PHY-2408988, and DMS-2204702 (P.G.K.).

\section*{Author contribution}
Y.M., C.C., P.G.K., and J.Y. conceptualized the work;
Y.M. and C.C. performed analytical calculations;
Y.M. conducted numerical simulations and experiments and drafted the manuscript.
All authors extensively contributed to the work and finalizing the manuscript.
J.Y. supervised the project.

\section*{Competing interest}
The authors declare no competing interests.

\endgroup

\clearpage

\end{document}